\documentclass{aa}  

\newcommand\dosingle[1]{#1}  \newcommand\dodouble[1]{ } 

\newcommand\nice[1]{#1}    \newcommand\subm[1]{}   
\subm{ \renewcommand\dosingle[1]{ }   \renewcommand\dodouble[1]{#1} }

\dodouble{ \documentclass[referee]{aa} }

\usepackage{graphicx}

\usepackage{amsfonts}
\usepackage{mathpi}

\usepackage{url} 

 \usepackage{hyperref}
\nice{ 
\providecommand{\eprint}[1]{\href{http://arxiv.org/abs/#1}{{\tt [arXiv:#1]}}}

\providecommand{\url}[1]{\href{#1}{#1}}

}

\providecommand{\adsurl}[1]{} 
\newcommand\SSS{Sect.~}

\nice{ \newcommand\mycaptionfont{\protect\footnotesize} }

\newcommand\kms{\,km\,s$^{-1}$}

\newcommand\gtapprox{\,\lower.6ex\hbox{$\buildrel >\over \sim$} \, }
\newcommand\ltapprox{\,\lower.6ex\hbox{$\buildrel <\over \sim$} \, }
\newcommand\propapprox{\,\lower.6ex\hbox{$\buildrel \propto\over \sim$} \, }

\newcommand\arcs{\ifmmode {'' }\else $'' $\fi}     
\newcommand\arcm{\ifmmode {' }\else $' $\fi}       
\newcommand\ddeg{\ifmmode^\circ\else$^\circ$\fi}    

\newcommand\frtoday{Le\space\number\day\space\ifcase\month\or
  janvier\or f\'evrier\or mars\or avril\or mai\or juin\or
  juillet\or ao\^ut\or septembre\or octobre\or novembre\or 
d\'ecembre\fi\space \number\year}

\newcommand\cqg{ClassQuantGra}   %
\newcommand\hGpc{\mbox{$h^{-1}$ Gpc}}
\newcommand\hMpc{\mbox{$h^{-1}$ Mpc}}

\newcommand\rSLS{r_{\mathrm{SLS}}}  
\newcommand\rC{R_{\mathrm{C}}}  
\newcommand\rhor{R_{\mathrm{hor}}} 

\newcommand\Omtot{\Omega_{\mathrm{tot}}}

\newcommand\ddotrtopo{\ddot{\bf{r}}_{\mbox{\rm topo}}}

\newcommand\PDSmath{\mathrm{S}^3\mathrm{/I}^*}

\newcommand\notea{^\mathrm{a}}
\newcommand\noteb{^\mathrm{b}}
\newcommand\notec{^\mathrm{c}}
\newcommand\noted{^\mathrm{d}}
\newcommand\notee{^\mathrm{e}}
\newcommand\notef{^\mathrm{f}}
\newcommand\noteg{^\mathrm{g}}
\newcommand\noteh{^\mathrm{h}}

\title{The residual gravity acceleration effect in the
  Poincar\'e dodecahedral space}

\author{Boudewijn F. Roukema
\and Piotr T. R\'o\.za\'nski
}

\institute{Toru\'n Centre for Astronomy, Nicolaus Copernicus University,
ul. Gagarina 11, 87-100 Toru\'n, Poland 
}

\date{\frtoday}

\titlerunning{Residual gravity in the Poincar\'e space}
\authorrunning{Roukema \& R\'o\.za\'nski}

\begin{document}


\newcommand\Nchainsmain{16}
\newcommand\Npergroup{four}

\abstract
{ In a flat space, it has been shown heuristically that the global
  topology of comoving space can affect the dynamics expected in the
  weak-field Newtonian limit, inducing a weak acceleration effect
  similar to dark energy.  }
{ Does a similar effect occur in the case of the Poincar\'e
  dodecahedral space, which is a candidate model of comoving space for solving
  the missing fluctuations problem observed in cosmic microwave
  background all-sky maps? Moreover, does the effect distinguish the
  Poincar\'e space from other well-proportioned spaces?  }
{ The acceleration effect in the Poincar\'e space $S^3/I^*$ is
  studied, using a massive particle and a nearby test particle of
  negligible mass.  Calculations are made in $S^3$ embedded in
  $\mathbb{R}^4$.  The weak-limit gravitational attraction on a test
  particle at distance $r$ is set to be $\propto [\rC \sin({r/\rC})]^{-2}$
  rather than $\propto r^{-2}$, where $\rC$ is the curvature radius,
  in order to satisfy Stokes' theorem. A finite
  particle horizon large enough to include the adjacent topological
  images of the massive particle is assumed.  The regular, flat,
  3-torus $T^3$ is re-examined, and two other well-proportioned
  spaces, the octahedral space $S^3/T^*$, and the truncated cube space
  $S^3/O^*$, are also studied.  }
{ The residual gravity effect occurs in all four cases. In a perfectly
  regular 3-torus of side length $L_a$, and in the octahedral and
  truncated cube spaces, the highest order term in the residual
  acceleration is the third-order term in the Taylor expansion 
  in powers of
  $r/L_a$ (3-torus), or $r/\rC$, respectively. However, the Poincar\'e
  dodecahedral space is unique among the four spaces. The third order
  cancels, leaving the fifth order term $\sim \pm 300 (r/\rC)^5$ as the
  most significant.  }
{ Not only are three of the four perfectly regular well-proportioned
  spaces better balanced than most other multiply connected spaces in
  terms of the residual gravity acceleration effect by a factor of
  about a million (setting $r/L_a = r/\rC \sim 10^{-3}$), but the
  fourth of these spaces is about ten thousand times better balanced than
  the other three. This is the Poincar\'e dodecahedral space.  Is this
  unique dynamical property of the Poincar\'e space a clue towards a
  theory of cosmic topology?  }


\maketitle

\dodouble{ \clearpage } 


\newcommand\tcoeff{
\begin{table}
\caption{\mycaptionfont 
Statistical characteristics of coefficients $a_i$ 
of the dominant ($i$-th order) term 
in the radial and orthogonal components of the residual acceleration $\ddot{r}$ in 
perfectly regular well-proportioned spaces, for approximately isotropic 
displacements $\mathbf{r}$.$\notea$
\label{t-coeff}}
$$\begin{array}{c c c r r r r} \hline  \hline
\mbox{space} \rule[-1.5ex]{0ex}{4.5ex}
& \mbox{term}\noteb
& \parallel/\perp\notec  
& \left<a_i\right>
& \sigma_{\left<a_i\right>}\noted
& \sigma_i\notee
& \gamma_i\notef
\\ \hline 
\rule{0ex}{2.5ex}    
\mbox{3-torus} & (r/L_a)^3 & \parallel &       0.00 &   0.00  & 6.11 &       0.58  \\ 
\mbox{3-torus} & (r/L_a)^3 & \perp &       6.39 & 0.00 & 6.82 & -0.42  \\ 
\mathrm{octahedral} & (r/\rC)^3 & \parallel &        0.01 &       0.01 &       3.18 &      -0.58  \\ 
\mathrm{octahedral} & (r/\rC)^3 & \perp &        3.33 &       0.00 &       1.26 &      -0.42  \\ 
\mathrm{tr. cube} & (r/\rC)^3 & \parallel &       -0.05 &       0.05 &      14.24 &       0.58  \\ 
\mathrm{tr. cube} & (r/\rC)^3 & \perp &       14.94 &       0.02 &       5.63 &      -0.42  \\ 
\mathrm{dodec/num}\noteg & (r/\rC)^5 & \parallel &       -0.30 &       0.53 &     288.29 &       0.75  \\ 
\mathrm{dodec/alg}\noteh & (r/\rC)^5 & \parallel &     0.00 &       0.01 &     288.26 &       0.74 \\ 
\mathrm{dodec/num}\noteg & (r/\rC)^5 & \perp &      286.33 &       0.22 &     121.65 &      -0.37  \\ 

\hline
\end{array}$$
\\
$\notea$ coefficients $a_i$ as defined in 
Eqs~(\protect\ref{e-define-a3-flat}),
(\protect\ref{e-define-a3-S3}), and
(\protect\ref{e-define-a5-S3}); these are approximately constant with respect to $r$;
the constant factor of $Gm/L_a^2$ for the $T^3$ model 
or $Gm/\rC^2$ for the other models has been ignored here; all values
shown are dimensionless \\
$\noteb$ dominant $i$-th power of displacement, as derived in this paper \\
$\notec$ radial $\parallel$ or orthogonal $\perp$ component \\
$\noted$ standard error in the mean
$\sigma_{\left<a_i\right>} =
\sigma_i/\sqrt{N-1}$ for $N \gg 1 $ test particles \\
$\notee$ sample standard deviation \\
$\notef$ sample skewness $\gamma_i = 
      \left<[(a_i- \left<a_i\right>)/\sigma_i]^3 \right>$ \\
$\noteg$ from 70-bit significand numerical calculations using 
      Eq.~(\protect\ref{e-residgrav-S3-Gamma}) \\
$\noteh$ using the algebraic expression in 
      Eq.~(\protect\ref{e-residgrav-Poincare-exact})
\end{table}
}  

\newcommand\tcoefftcube{
\begin{table}
\caption{\mycaptionfont 
Components of the residual acceleration $\ddot{r}$, 
as for Table~\protect\ref{t-coeff},
for the truncated cube space,
separated by fundamental domain face shape.
\label{t-coeff-tcube}}
$$\begin{array}{c c c r r r r} \hline  \hline
\mbox{space} \rule[-1.5ex]{0ex}{4.5ex}
& \mbox{term}
& \parallel/\perp
& \left<a_i\right>
& \sigma_{\left<a_i\right>}
& \sigma_i
& \gamma_i
\\ \hline 
\rule{0ex}{2.5ex}    
\mathrm{octagonal}\notea & (r/\rC)^3 & \parallel &       -0.00 &       0.04 &      17.46 &       0.58  \\ 
\mathrm{octagonal} & (r/\rC)^3 & \perp &       18.27 &       0.02 &       6.89 &      -0.43  \\ 
\mathrm{triangular}\noteb & (r/\rC)^3 & \parallel &        0.00 &       0.01 &       3.19 &      -0.58  \\ 
\mathrm{triangular} & (r/\rC)^3 & \perp &        3.33 &       0.00 &       1.26 &      -0.43  \\ 
\hline
\end{array}$$
\\
$\notea$ contribution from the octagonal (truncated square) faces \\
$\noteb$ contribution from the triangular faces
\end{table}
}  

\newcommand\fTthreeAthree{
\begin{figure}
\centering 
\includegraphics[width=8cm]{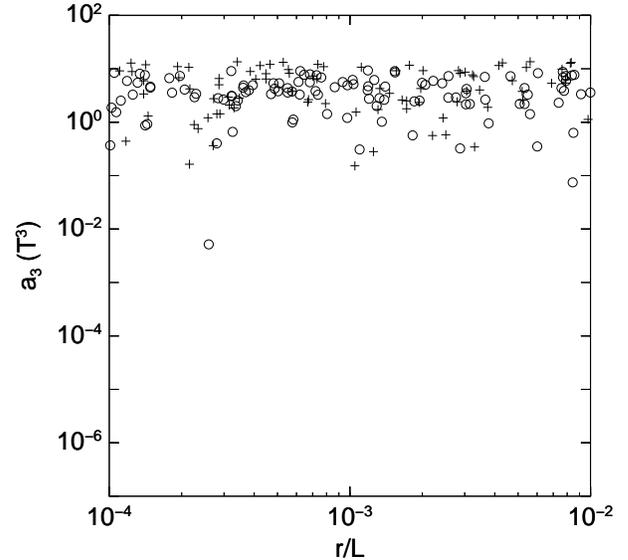}
\caption[]{ \mycaptionfont
Residual acceleration (radial component)
induced on a  test particle 
near a massive particle such as a cluster of galaxies 
by the six adjacent
topological images of the massive particle,
in a $T^3$ universe of three exactly comoving equal side lengths $L_a$.
The accelerations $a_3(r)$ are normalised to be constant if the dominant term is
third order in $r/L_a$ [see Eq.~(\protect\ref{e-define-a3-flat})] and are shown
against distance $r < 0.01 L_a$ from the massive particle, 
where $L=3000{\hMpc}.$ Positive accelerations are shown by plus symbols; 
negative accelerations are shown by circles.
}
\label{f-T3-a3}
\end{figure} 
} 

\newcommand\fdodecAfive{
\begin{figure}
\centering 
\includegraphics[width=8cm]{dodec_a5l.ps}
\caption[]{ \mycaptionfont
Residual accelerations for the Poincar\'e dodecahedral space $\PDSmath$,
as for Fig.~\protect\ref{f-dodec-a3}, 
but normalised to be constant if the {\em fifth}
order term dominates, i.e. $a_5$ [see Eq.(~\protect\ref{e-define-a5-S3})]
is shown. The
vertical scale differs from the previous
figures.
}
\label{f-dodec-a5}
\end{figure} 
} 

\newcommand\fdodecAthree{
\begin{figure}
\centering 
\includegraphics[width=8cm]{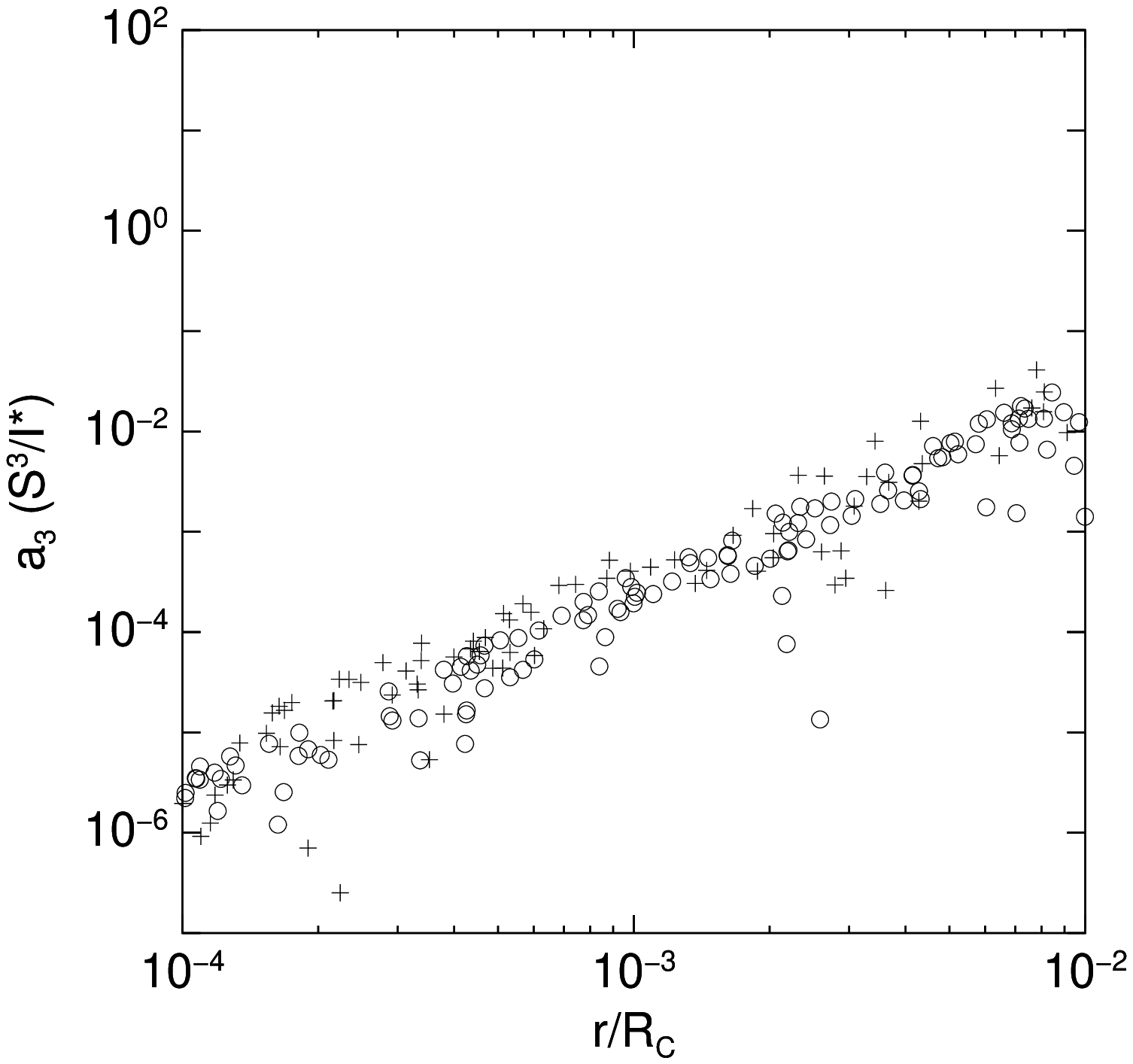}
\caption[]{ \mycaptionfont
Residual acceleration (radial component),
as for Fig.~\protect\ref{f-octa-a3}, 
for the Poincar\'e dodecahedral space $\PDSmath$.
}
\label{f-dodec-a3}
\end{figure} 
} 

\newcommand\foctaAthree{
\begin{figure}
\centering 
\includegraphics[width=8cm]{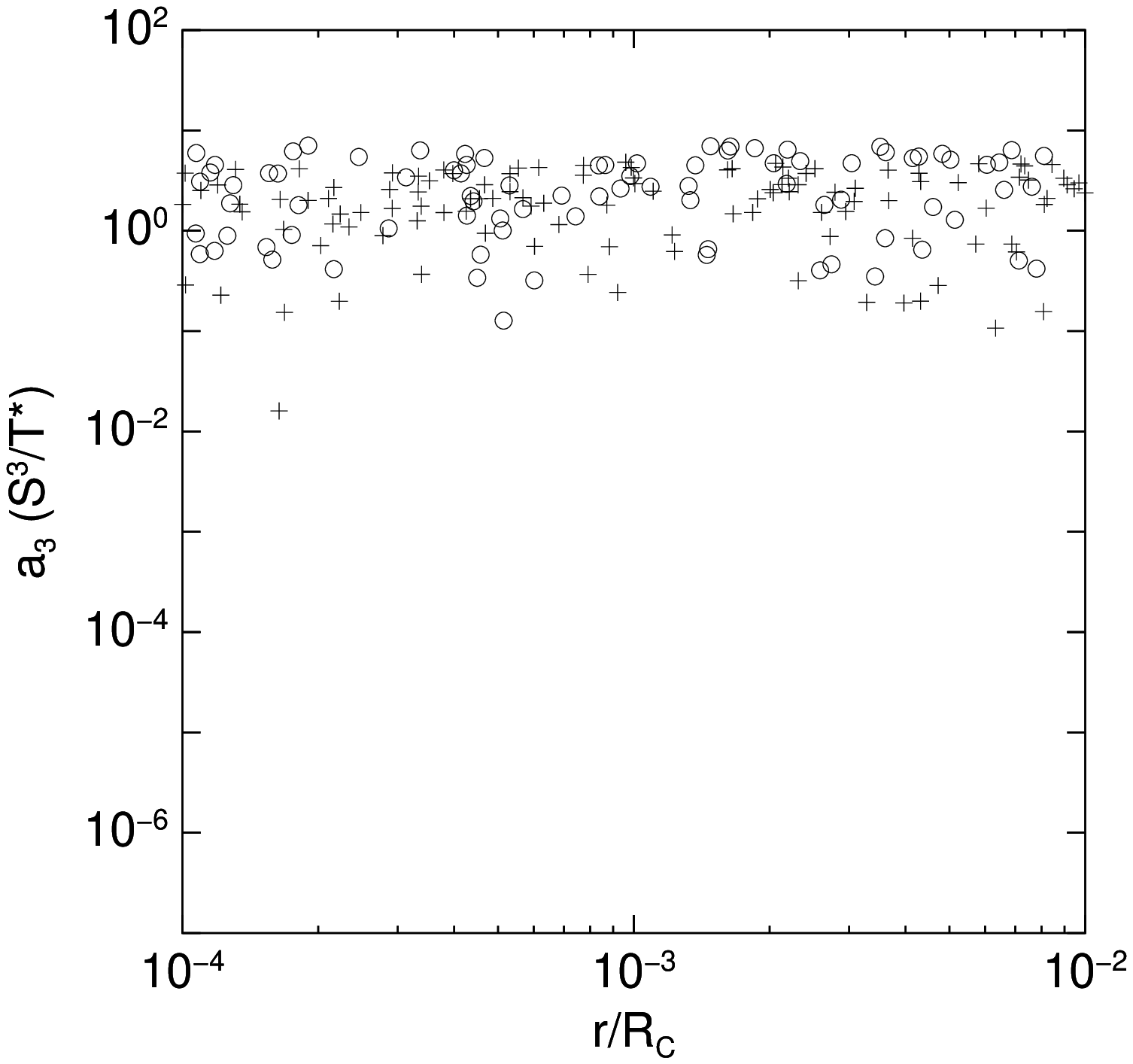}
\caption[]{ \mycaptionfont
Residual acceleration (radial component)
induced on a  test particle 
near a massive particle by the eight adjacent
topological images of the massive particle,
in a universe whose 3-manifold of comoving space is 
the octahedral space $S^3/T^*$.
The accelerations  $a_3(r)$ 
are normalised to be constant if the dominant term is
third order in $r/L_a$
[see Eq.~(\protect\ref{e-define-a3-S3})] and are shown
against distance $r < 250 {\hMpc} = 0.01 \rC $ from the massive particle. 
Positive accelerations are shown by plus symbols; 
negative accelerations are shown by circles.
}
\label{f-octa-a3}
\end{figure} 
} 

\newcommand\ftcubeAthree{
\begin{figure}
\centering 
\includegraphics[width=8cm]{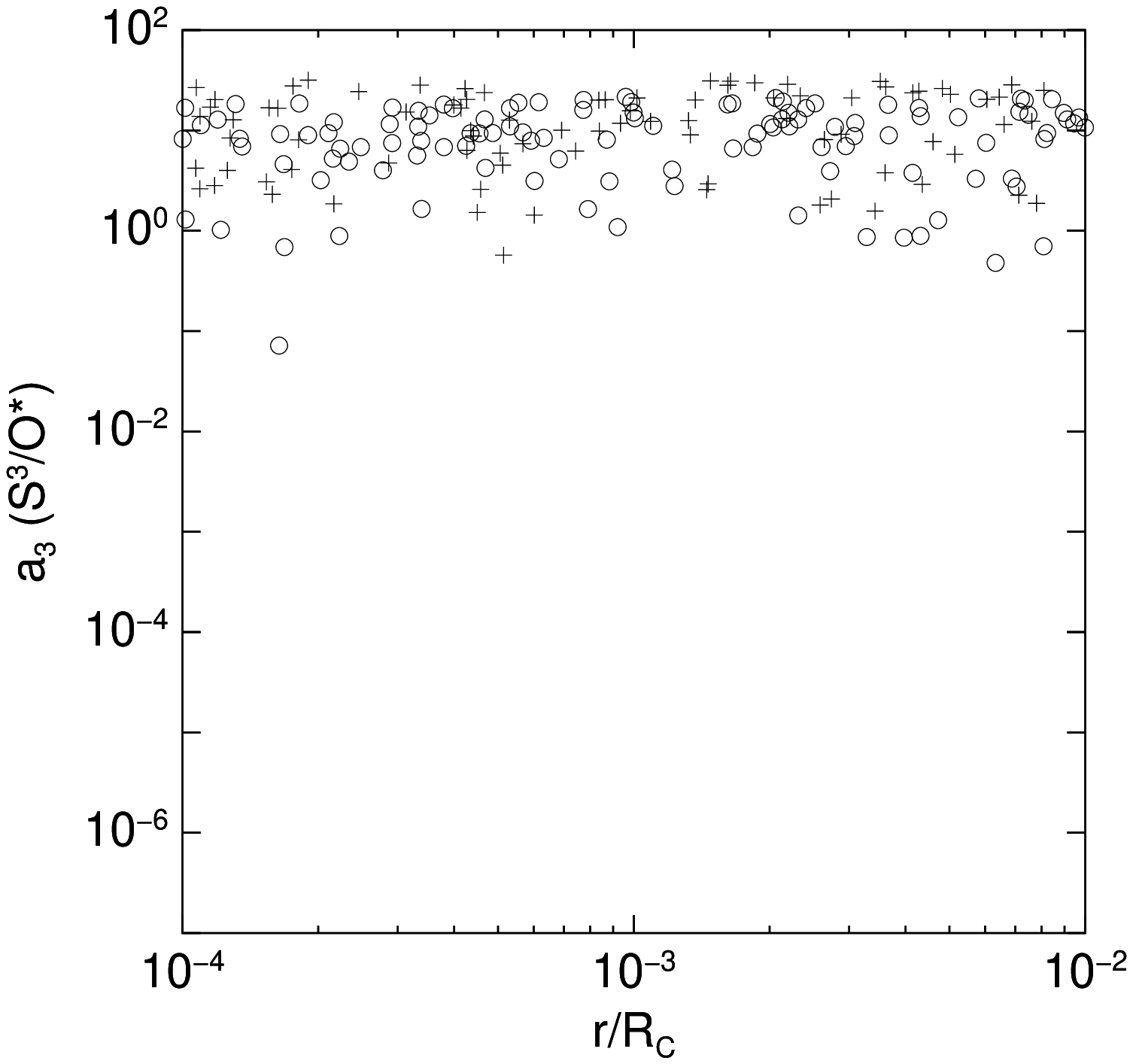}
\caption[]{ \mycaptionfont
As for Fig.~\protect\ref{f-octa-a3}, 
for the truncated cube space
$S^3/O^*$.
}
\label{f-tcube-a3}
\end{figure} 
} 

\newcommand\fdodecAthreeperp{
\begin{figure}
\centering 
\includegraphics[width=8cm]{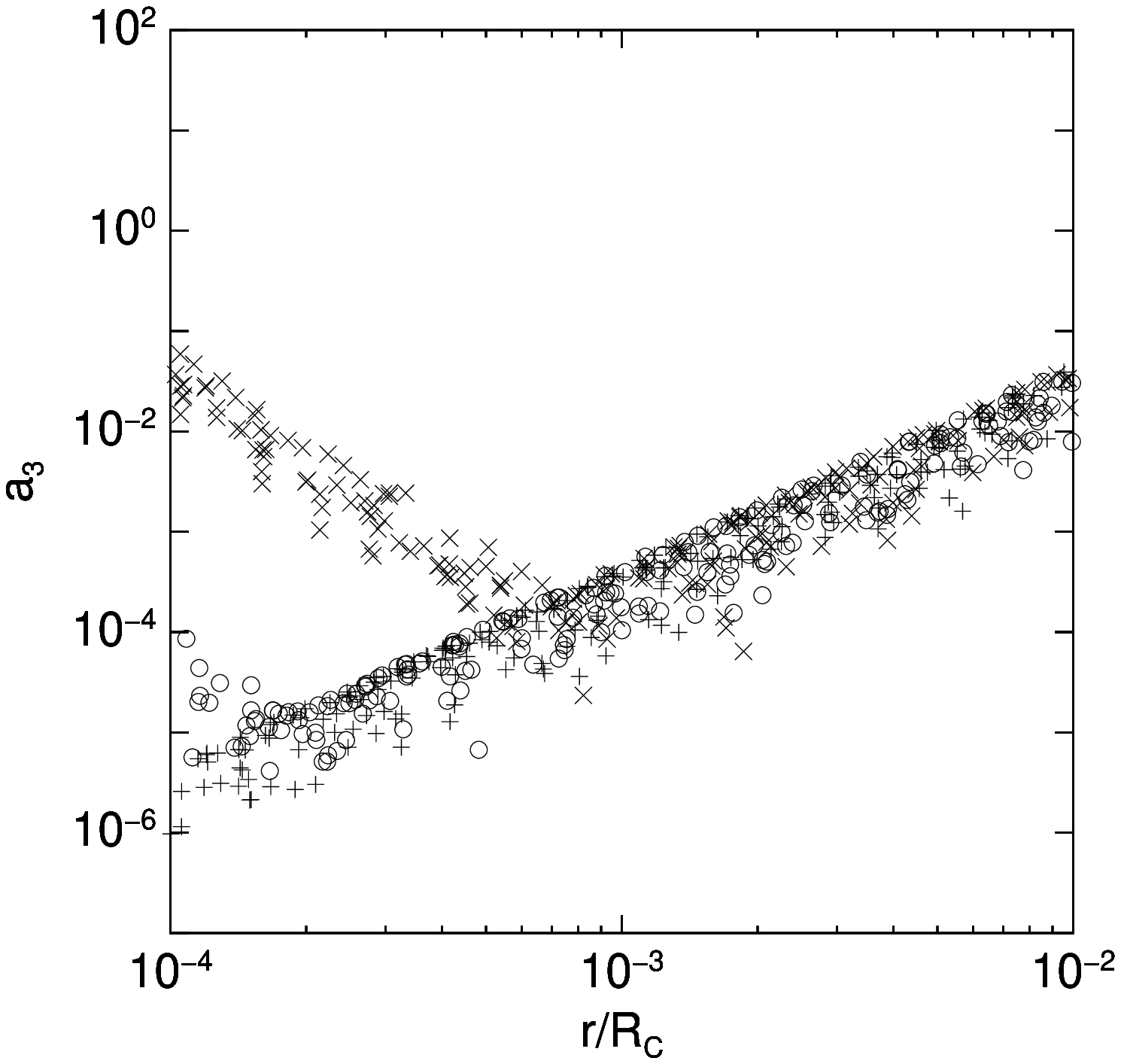}
\caption[]{ \mycaptionfont
Orthogonal residual acceleration 
for the Poincar\'e dodecahedral space $\PDSmath$,
as for Fig.~\protect\ref{f-dodec-a3}, shown 
using 50-bit ($\times$), 
60-bit ($\circ$) and 70-bit ($+$)
precision in the significands of the floating-point
numbers.
}
\label{f-dodec-a3-perp}
\end{figure} 
} 


\section{Introduction}  \label{s-intro}

It has been shown that for zero curvature,
the global topology of comoving space can affect the dynamics
expected in the weak-field Newtonian limit \nocite{RBBSJ06}({Roukema} {et~al.} 2007), in contrast to what
was previously thought. In particular, a 
test particle of negligible mass
near a massive particle (such as a cluster of galaxies
dominated by its dark matter halo)
has unequal attractions 
to the nearest topological images of the massive particle in opposite
directions, since its
position is asymmetrically offset from the massive particle. This leaves a residual
acceleration effect that is qualititatively 
similar to that of dark energy. For realistic physical scales at the 
present epoch, the
effect was estimated to be about $10^{-9}$ times weaker than the observed
cosmological constant.

The heuristic calculations presented in \nocite{RBBSJ06}{Roukema} {et~al.} (2007) only
considered the flat spaces ${T} \times \mathbb{R}^2$ and
${T}^3$. 
However, there has been interest in the 
Poincar\'e dodecahedral space,
$S^3/I^*$, being a candidate model for comoving space that
explains the ``missing fluctuations problem'' 
tentatively observed in the
COsmic Microwave Background (COBE) all-sky maps
and confirmed in the Wilkinson Microwave Anisotropy Probe (WMAP) all-sky maps
\nocite{WMAPSpergel,LumNat03,RLCMB04,Aurich2005a,Aurich2005b,Gundermann2005,KeyCSS06,NJ07,Caillerie07,LR08,RBSG08,RBG08}({Spergel} {et~al.} 2003; {Luminet} {et~al.} 2003; {Roukema} {et~al.} 2004; {Aurich} {et~al.} 2005a, 2005b; {Gundermann} 2005; {Key} {et~al.} 2007; {Niarchou} \& {Jaffe} 2007; {Caillerie} {et~al.} 2007; {Lew} \& {Roukema} 2008; {Roukema} {et~al.} 2008a, 2008b).
Does a similar residual gravity effect occur in the case of
positively curved space, in particular, in the Poincar\'e 
space?  

An additional question of interest is whether or not the residual
gravity effect might distinguish the Poincar\'e space from other
well-proportioned spaces \nocite{WeeksWellProp04}({Weeks} {et~al.} 2004). The Poincar\'e space
is presently preferred to the octahedral space ($S^3/T^*$) and the
truncated cube space ($S^3/O^*$) because of empirical constraints on 
curvature. For a
total density of $1.02 \gtapprox \Omtot \gtapprox 1.01$, the latter
two spaces have larger fundamental domains than the Poincar\'e space and so
have difficulty explaining the missing fluctuations problem. However,
this is an empirical argument with no theoretical motivation.  Could
there be any dynamical arguments either favouring or disfavouring
the Poincar\'e space? Some steps have been taken towards what might
develop into a quantum cosmology theory of cosmic topology, using various
notions of distance between different manifolds \nocite{Masafumi96,CarlipSurya04}({Masafumi} 1996; {Anderson} {et~al.} 2004) 
and analysis of topology change in quantum gravity \nocite{DowS98}(e.g.,  {Dowker} \& {Surya} 1998).
The residual gravity effect might contribute an additional criterion
for comparing different manifolds.

Here, the residual gravity effect in 
the octahedral space, the truncated cube space,
and the Poincar\'e space is studied
by considering the dynamics of a negligible-mass test particle near a 
massive particle.
As in \nocite{RBBSJ06}{Roukema} {et~al.} (2007), this approach can be considered as an heuristic model
for a positively curved space that is homogeneous except for a small
neighbourhood around one point, in which a positive density fluctuation 
of matter has collapsed into
a high-density, nearly point-like object, in excess of the underlying
homogeneous density distribution.
To satisfy Stokes' theorem, 
the weak-limit gravitational attraction in the three spherical spaces
is set to be proportional to 
\begin{equation}
\left[ \rC \sin\left(\frac{r}{\rC}\right) \right]^{-2}
\label{e-sph-grav}
\end{equation} 
rather than $r^{-2}$ 
{(see \SSS\ref{s-sph-grav}),}
where $\rC$ is the curvature radius.
Numerical calculations are made in the universal covering space $S^3$
embedded in $\mathbb{R}^4$ for convenience. Accelerations are calculated 
in the tangent 3-space at the position of the test particle,
represented as 4-vectors in $\mathbb{R}^4$.
The $T^3$ model discussed in \nocite{RBBSJ06}{Roukema} {et~al.} (2007) is revisited, using
the covering space $\mathbb{R}^3$.


The method is described in more detail in \SSS\ref{s-method}.
Residual accelerations estimated to the third order in the Taylor
expansion of the fractional displacement are presented for the 3-torus
in \SSS\ref{s-T3-3rdorder}. Residual accelerations to fifth order for
individual pairs of opposite images in the spherical spaces are presented
analytically in \SSS\ref{s-results-onepair}. Residual accelerations to
fifth order for the full set of adjacent\footnote{``Adjacent'' is used
here to refer to images in copies of the fundamental domain that share
a face with the ``original'' copy of the fundamental domain.} 
images in the spherical spaces are presented
numerically in \SSS\ref{s-results-wellprop}.  
For the Poincar\'e space, an analytical
derivation of the residual acceleration 
to fifth order is also given in \SSS\ref{s-results-analytical}. 
A statistical summary is presented in 
\SSS\ref{s-results-stats}.

Discussion and
conclusions are given in Sects.~\ref{s-disc} and \ref{s-conclu}
respectively.  Discussion of spherical, multiply connected spaces is
available in \nocite{Weeks2001}{Weeks} (2001), \nocite{GausSph01}{Gausmann} {et~al.} (2001), \nocite{LehSph02}{Lehoucq} {et~al.} (2002),
and \nocite{RiazSph03}{Riazuelo} {et~al.} (2004), and we refer the reader to references therein for
introductions to cosmic topology. Distances are calculated in a spatial section
at constant cosmological time 
in the universal covering space ($S^3$ for the spherical spaces,
$\mathbb{R}^3$ for $T^3$), by default as ``physical''
distances, i.e., $r = a\chi$, where $a(t)$ is the scale factor 
and $\chi$ is a comoving distance 
(i.e. $\chi$ is a ``proper
distance'' at the present epoch, \nocite{Wein72}({Weinberg} 1972), equivalent to
``conformal time'' if $c=1$). Distance units are normally presented here in $\hMpc$ 
or $\hGpc$, 
where the Hubble constant is written $H_0 \equiv 100 h$\kms/Mpc.  
The fundamental domains of all the spaces are assumed to be
perfectly regular. 
{The Newtonian gravitational constant is written as $G$.}


\section{Method} \label{s-method}

Although dynamics in the Poincar\'e dodecahedral space $S^3/I^*$ 
could, in principle, be studied by applying boundary conditions that identify opposite faces of a dodecahedral fundamental domain of a positive 
curvature radius $\rC$ with one another, this would probably be extremely
complicated for both analytical and numerical calculations. It is 
simpler to work in the universal covering space $S^3$ represented as 
a subspace of $\mathbb{R}^4$.


As described in Sect.~2.1 of \nocite{RBBSJ06}{Roukema} {et~al.} (2007), several assumptions are
required for this heuristic approach. Here, the corresponding assumptions 
are as follows:
\begin{list}{(\arabic{enumi})}{\usecounter{enumi}}
\item the flat Newtonian approximation of gravity is replaced by the
  equivalent in positively curved space, as described in Eq.~(\ref{e-sph-grav})
  above;
\item[(2), (3)] the covering space is $S^3$, which is not flat; 
  \addtocounter{enumi}{2}
\item a finite particle horizon $\rhor$ that is just large enough to include
  all adjacent topological images is assumed;
\item[(5), (6)] 
identical to those in \nocite{RBBSJ06}{Roukema} {et~al.} (2007): the metric
  is assumed to be that for a perfectly homogeneous model of the 
  same curvature, except that the distant, multiple copies 
  of the local massive particle are considered to provide a
  contribution to the local gravitational potential that may
  not fully cancel;
  \addtocounter{enumi}{2}
\item[(7), (8)] are not needed here since the one-body problem in $S^3$ 
  (\SSS3.1.1, \nocite{RBBSJ06}{Roukema} {et~al.}) is divergent and only briefly
  mentioned in \SSS\ref{s-sph-grav}.
  \addtocounter{enumi}{2}  
\end{list}

\subsection{Weak-limit gravity and divergences} \label{s-sph-grav}

As stated in Eq.(\ref{e-sph-grav}), to satisfy Stokes'
theorem, the weak-limit gravitational attraction 
{towards a single massive particle}
is set to be
proportional to $[\rC \sin({r}/{\rC}) ]^{-2}$ rather than
$r^{-2}$. {This can be understood as follows.  

  Let $S^3$ be represented by a spherical coordinate system centred on
  the massive particle, so that the
  Friedmann-Lema\^{\i}tre-Robertson-Walker (FLRW) line element is written
  \begin{eqnarray}
    \mathrm{d}s^2 
    &=& -c^2 \mathrm{d}t^2  + 
    \mathrm{d}r^2 + [\rC \sin(r/\rC)]^2 (
            \mathrm{d}\theta^2 + \cos^2\theta\, \mathrm{d}\phi^2 ),
            \nonumber \\
  \end{eqnarray}
  where the scale factor $a(t)$ is subsumed into $r$ and
  $\rC$, since it is not needed explicitly.  We can write the weak-limit
  gravitational acceleration towards the massive particle as the vector
  field
  \begin{equation}
    f(r)\; \hat{\mathbf{r}}
  \end{equation}
  by symmetry, where 
  $\hat{\mathbf{r}}$ is the unit radial vector in the tangent space
  at a radial ``physical'' distance $r$ from the origin along a spatial
  geodesic, and
  \begin{equation}
  \lim_{r\rightarrow 0^+}  \frac{ -f(r)\; r^{2} }{Gm} 
  = 1
  \label{e-newt-limit}
  \end{equation}
  in order to agree with the flat-space Newtonian limit towards the
  massive particle of mass $m$.  Let $V$ be the
  interior of a 2-sphere of radius $r$ centred at the massive particle. 
  Applying Stokes' theorem to $f(r)\; \hat{\mathbf{r}}$ on
  $V$, which is a sub-manifold of $S^3$ with boundary $\partial V$, gives
  \begin{eqnarray}
    \int_V \mathrm{div} [f(r')\; \hat{\mathbf{r}}] \, \mathrm{d}V 
    &=& 
    \int_{\partial V} f(r')\; \hat{\mathbf{r}}  \,.\, \mathrm{d}\mathbf{S} 
      \nonumber \\
      &=&     \int_{\partial V} f(r') [\rC \sin(r'/\rC)]^2 \; 
      \mathrm{d}\theta \;  
      \cos\theta \mathrm{d}\phi ,
      \nonumber \\
      &=&     f(r) [\rC \sin(r/\rC)]^2 \; 
      \int_{\partial V} \mathrm{d}\theta \;  
      \cos\theta \mathrm{d}\phi ,
      \nonumber \\
      &=& 4\pi [\rC \sin(r/\rC)]^2 \; f(r),
      \label{e-stokes-howto}
  \end{eqnarray}
  using the FLRW metric.
  Since the only source of the vector field $f(r)\; \hat{\mathbf{r}}$
  is the single massive particle, the left-hand side of this equation
  equals its limit for arbitrarily small (non-zero) volumes, i.e.
  \begin{eqnarray}
    \int_V \mathrm{div} [f(r')\; \hat{\mathbf{r}}]  \, \mathrm{d}V  
    &=&
    \lim_{r\rightarrow 0^+} \int_V \mathrm{div} [f(r')\; \hat{\mathbf{r}}]  \, \mathrm{d}V  
    \nonumber \\
    &=&   
    \lim_{r\rightarrow 0^+} 4\pi [\rC \sin(r/\rC)]^2 \; f(r) 
    \nonumber \\
    &=& -4\pi G m,
    \label{e-stokes-flat}
  \end{eqnarray}
  using Eqs~(\ref{e-stokes-howto}) and (\ref{e-newt-limit})
  respectively.  Equating the right-hand sides of
  Eqs~(\ref{e-stokes-howto}) and (\ref{e-stokes-flat}) gives the weak
  limit acceleration induced by a single massive particle in a
  positively curved space of radius $\rC$
  \begin{equation}
    \ddot{\mathbf{r}} = f(r) \; \hat{\mathbf{r}} = -\frac{G m \;
      \hat{\mathbf{r}} } {[\rC \sin(r/\rC)]^2}.
         \label{e-f-final}
  \end{equation}
For} instantaneous transmission of gravitational signals in an
infinitely old, static, simply connected space,
this implies two divergence problems.

Firstly, a single massive point particle of mass $m$ 
in a perfectly uniform and otherwise empty 
$S^3$ yields an infinite repulsive force at the antipode of the massive particle,
since $\sin(\pi) = 0$. More generally, an infinite repulsive (attractive) force
exists for gravitational signals that have travelled 
odd (even) values of $j$ times the half-circumference $\pi \rC$ 
from the massive object, i.e. $j/2$ times around the whole space,
where $j \in \mathbb{Z}$.

Secondly, even if we ignore the model of a zero-size point particle as an
idealised fiction, a problem remains for a 
negligible-mass test particle near a massive
particle.
The test particle 
also experiences accelerations from signals that have travelled $j$ 
times around $S^3$ in the two different directions along the great circle
passing through the test particle and the massive particle, 
where $j \in \mathbb{Z}$.
This second divergence problem is similar to the divergence problem
in flat, simply connected, infinitely sized, infinitely old, 
Newtonian space with instantaneous 
transmission of the gravitational signal. The total acceleration 
towards the massive particle would be
\begin{eqnarray}
  \ddot{r}_{\mathrm{S}^3} &=& 
  \sum_{j=0}^\infty  G m \left( \rC \sin \frac{r+ 2j\pi \rC}{\rC}  \right)^{-2} 
   \nonumber \\
  && - \sum_{j=-\infty}^{-1} G m \left( \rC \sin \frac{r+ 2j\pi \rC}{\rC}  \right)^{-2} 
\nonumber \\
   &=& 
  G m \left( \rC \sin \frac{r}{\rC}  \right)^{-2} 
  \left( \sum_{j=0}^\infty  1  - 
  \sum_{j=-\infty}^{-1}  1 \right), 
\label{e-S3-sum-divergence}
\end{eqnarray}
which is clearly divergent.

The former divergence, i.e., the $j\pi \rC$ singularity related 
to the zero size of the massive point particle and the nature
of positive curvature, is clearly unphysical if we are interested in 
a smoothing length on the scale of a galaxy cluster. 
%
%
The latter divergence, i.e. that in Eq.~(\ref{e-S3-sum-divergence}), 
is removed by assumption (4) above. Assumption (4) also removes the
$j\pi \rC$ divergence.
A finite particle horizon $\rhor$ 
just large enough to include the adjacent topological images is necessary
for a residual gravity effect to occur, but it does not need to
be as large as $\pi \rC$.
The adjacent topological images in 
$S^3/T^*$,
$S^3/O^*$, and
$S^3/I^*$  are at $(\pi/3) \rC$, $(\pi/4) \rC$ and $(\pi/3) \rC$, and $(\pi/5) \rC$ respectively.
Moreover, 
observationally realistic estimates of the
total density parameter\footnote{For example,
$\Omtot = 1.010_{-0.009}^{+0.016}$, from WMAP 3-year and Hubble Space 
Telescope $H_0$ key project data,
$\Omtot = 1.015_{-0.016}^{+0.020}$, from WMAP 3-year data and Supernova Legacy 
Survey supernovae Type Ia data \nocite{WMAPSpergel06}({Spergel} {et~al.} 2007);
$\Omtot = 1.03^{+0.04}_{-0.06}$, from 
WMAP 3-year and Arcminute Cosmology Bolometer Array Receiver
ACBAR data \nocite{ACBAR08}({Reichardt} {et~al.} 2008).} 
are consistent with the range that is empirically interesting
for the Poincar\'e space,
$1.02 \gtapprox \Omtot \gtapprox 1.01$.
The range $1.02 \gtapprox \Omtot \gtapprox 1.01$ corresponds to a curvature radius of
$20 \hGpc \ltapprox \rC \ltapprox 30 \hGpc$ respectively. 
Hence, for the three well-proportioned, spherical spaces,
the appropriate horizon distances for the arrival of gravitational
signals from the adjacent topological images are in the range
$10 \hGpc \ltapprox \rhor \ltapprox 30 \hGpc$, i.e. up to a few times 
the distance to the surface of last scattering $\rSLS \approx 10\hGpc$.
As noted in \nocite{RBBSJ06}{Roukema} {et~al.} (2007), a moderate amount of inflation in the early Universe could be
one way of satisfying assumption (4).

\subsection{Residual gravity} \label{s-meth-resgrav}

The weak-limit residual acceleration on a test particle at 
$\mathbf{r}$ near the massive particle of mass $m$  
at $\mathbf{0}$ 
can be written 
\begin{eqnarray}
  \ddot{\mathbf{r}} &=& 
  \sum_{j=1,d_j \equiv a\chi[\mathbf{r},g_j(\mathbf{0})]  }^{j=N}  
    G m \left( \rC \sin \frac{d_j}{\rC}  \right)^{-2} 
  \hat{\mathbf{r}}_j, 
\label{e-residgrav-S3-Gamma}
\end{eqnarray}
where $g_j = 1, \ldots, N$ are the $N=8$, 14, or 12 holonomies that
map an arbitrary object in 
$S^3/T^*$, $S^3/O^*$, or $\PDSmath$ respectively to its adjacent
topological images, 
$a$ is the scale factor,
$\chi(\mathbf{a},\mathbf{b})$ is the comoving distance
between two positions $\mathbf{a},\mathbf{b}$ in the covering space $S^3$,
$\hat{\mathbf{r}}_j$ is the unit vector in the tangent space to
$S^3$ at $\mathbf{r}$ pointing along the geodesic towards
$g_j(\mathbf{0})$, and $\ddot{\mathbf{r}}$ is the
acceleration vector in the same tangent space.  The appendices in
\nocite{GausSph01}{Gausmann} {et~al.} (2001) provide a convenient procedure for calculating this,
where a 3-manifold $S^3/\Gamma$ and holonomy group $\Gamma$
are modelled as a tiling and isometry group of $\mathbb{R}^4$ 
using quaternions and 4-dimensional matrices.  The tangent space at $\mathbf{r}$ is a
3-plane, modelled using vectors in $\mathbb{R}^4$. The unit tangent vector
towards the $j$-th topological image of the massive particle can be written
\begin{equation}
  \hat{\mathbf{r}}_j =  \frac{ 
    g_j(\mathbf{0}) -  ( \widehat{g_j(\mathbf{0})} . \widehat{\mathbf{r}} ) \mathbf{r} }{
    || g_j(\mathbf{0}) -  ( \widehat{g_j(\mathbf{0})} . \widehat{\mathbf{r}} ) \mathbf{r} || }
\label{e-rjhat-howto}
\end{equation}
where $\widehat{\mathbf{x}} \equiv \mathbf{x} / || \mathbf{x} || $.  A useful
identity is
\begin{eqnarray}
  \left( \sin \frac{d_j}{\rC}  \right)^{-2}  &=&
    \left( \sin \left\{ \mathrm{acos}
      \left[  \widehat{g_j(\mathbf{0})} . \widehat{\mathbf{r}}\right] \right\} \right)^{-2} 
      \nonumber \\
 &=&    \left\{ 1-
      \left[  \widehat{g_j(\mathbf{0})} . \widehat{\mathbf{r}}\right]^2 \right\}^{-1} .
\label{e-sinacos-simplify}
\end{eqnarray}
The curvature is
set to $\rC \equiv 25\hGpc$ below unless stated otherwise.
The horizon $\rhor$ is set sufficiently
large to include the full set of adjacent images of the massive particle,  
i.e., 
values of $\rhor = 1.1 (\pi/3) \rC$, $1.1 (\pi/3) \rC$, and  $1.1 (\pi/5) \rC$ are 
used in Eq.~(\ref{e-residgrav-S3-Gamma}) for $S^3/T^*$, $S^3/O^*$, and $\PDSmath$,
respectively.

\subsection{Numerical precision} \label{s-meth-numerical}


For the full set of adjacent topological images in the spherical
spaces, Eq.~(\ref{e-residgrav-S3-Gamma}) is evaluated 
primarily by a numerical approach, using double precision
floating-point operations where sufficient, and arbitrary precision
floating-point operations where necessary.  Typical scales of
interest, i.e., for typical test particles in a void of large-scale 
structure, are a few tens of megaparsecs.  
This gives $10^{-4} < r/\rC < 10^{-2}$ as a useful
domain 
for finding the highest terms in the Taylor expansion of the residual gravity
for $r \ll \rC$, where $\rC = 25\hGpc$ as stated above.  
By
setting $\rC = G = m = 1$ internally in numerical calculations, nearly equal
and opposite accelerations from nearly opposed topological images are
each of approximately unity order, the fourth and fifth order
terms at the lower limit $r/\rC = 10^{-4}$ should therefore be above the
numerical noise limit, if the calculating precision is well below $\sim
10^{-16}$ and $10^{-20}$, respectively. These would require precision
in the significand\footnote{The IEEE 754-2008 standard recommends the
  term ``significand'' rather than ``mantissa''.}  well above $\log_2
10^{16} \approx 53$ and $\log_2 10^{20} \approx 66$ bits respectively.
According to the IEEE 754-1985 standard, double-precision
floating-point numbers have 53-bit precision in the significand
(including one implicit bit). This would at best provide just one
bit of information for a fourth order term and no information for 
a fifth order term for $r/\rC = 10^{-4}$. For this reason, 
an arbitrary precision library is used here to examine higher
order terms.

Analytical calculations are also made for the Poincar\'e space.
We also revisit the $T^3$ calculation made in \SSS3.2 of \nocite{RBBSJ06}{Roukema} {et~al.} (2007), in
order to consider higher order terms. The first order term was
found to cancel for a perfectly regular $T^3$ model.

\section{Results}  \label{s-results}

\fTthreeAthree

\subsection{$T^3$ revisited} \label{s-T3-3rdorder}
In \SSS3.2 of \nocite{RBBSJ06}{Roukema} {et~al.} (2007), an analytical estimate of the
acceleration from all six adjacent images in a $T^3$ model was shown
to be zero to first order when the three fundamental lengths of the
model are exactly equal. However, the lower scatterings of points in
Figs.~7 and 8 of that paper do not represent numerical error.
Recalculation of Eqs~(11) to (15) 
of \nocite{RBBSJ06}{Roukema} {et~al.} (2007)
to higher order shows that for
perfectly equal fundamental lengths $L_{\mathrm{a}} = L_{\mathrm{e}} =
L_{\mathrm{u}}$, higher order terms in Eq.~(15) do not all cancel.
The lowest order non-cancelling terms are the third order terms
\begin{eqnarray}
&&(\ddotrtopo)_{(i,j,k)\in\{(\pm1,0,0),(0,\pm1,0),(0,0,\pm1)\}}
\nonumber \\
&=& 
\frac{7Gm}{L_a^2} [ 
  2\epsilon_x^3 - 3\epsilon_x(\epsilon_y^2 + \epsilon_z^2), 
  2\epsilon_y^3 - 3\epsilon_y(\epsilon_x^2 + \epsilon_z^2),  \nonumber \\
 && \quad\quad\quad 2\epsilon_z^3 - 3\epsilon_z(\epsilon_x^2 + \epsilon_y^2) ],
\label{e-test-object-T3-3rdorder}
\end{eqnarray}
where the massive particle is at the origin $(0,0,0)$ in the covering
space $\mathbb{R}^3$ and the test particle is slightly displaced at
$(x=\epsilon_x L_a,y=\epsilon_y L_a,z=\epsilon_z L_a)$.
For test particles distributed isotropically around the massive particle,
numerical integration of these third order terms
gives the distribution of accelerations. This distribution has
a mean and standard deviation radial acceleration of
\begin{equation} 
\ddot{r} \approx (0.00 \pm 6.11) 
\frac{Gm}{L_a^2} \left(\frac{r}{L_a}\right)^3 
\label{e-accel-T3}
\end{equation}
and a skewness $\approx 0.58 (Gm/L_a^2)(r/L_a)^3$,
where $r \equiv \sqrt{x^2 + y^2 +z^2}$. 
This expression is consistent with the 
lower scatterings of points in  Figs.~7 and 8 of \nocite{RBBSJ06}{Roukema} {et~al.} (2007).
Here, Fig.~\ref{f-T3-a3} shows a similar numerical calculation,
using the radial component of the 
sum of $1/r^2$ accelerations for the eight adjacent images,
calculated directly from Eq.~(10) of \nocite{RBBSJ06}{Roukema} {et~al.} (2007) without
using the Taylor expansion.
The acceleration is shown as 
\begin{equation}
a_3(r) \equiv \frac{\ddot{r}}{Gm/L_a^2} \left(\frac{r}{L_a}\right)^{-3}
\label{e-define-a3-flat}
\end{equation}
for convenience.

The analytical and numerical calculations are clearly consistent 
in showing that in the case of perfectly equal fundamental side lengths
of a $T^3$ model, there {\em is} indeed a residual topological gravity effect.
The residual 
acceleration is anisotropic, depending on the relation between the test particle's 
displacement from the massive particle and the orientation of the fundamental
directions. The positive skewness implies that even though the mean radial
acceleration is close to zero, the mode and median are negative, i.e., test
particles placed in random directions relative to the massive particle
are more likely to be subject to a 
radial deceleration rather than a radial acceleration.
The amplitude of the effect is third order, i.e., about 
$ (r/L_a)^2$ times smaller than the residual effect that occurs for 
slightly unequal fundamental lengths.
If an upper estimate for $r/L_a$ at the present epoch is used, 
then this is a factor of about a million.

\subsection{One pair of opposite topological images in $S^3/\Gamma$} 
\label{s-results-onepair}

Before considering the full effect from a layer of topological images
in the spherical cases,
let us first
consider the effect of just one pair of opposite topological
images of the ``local'' massive particle. 
This is somewhat similar to the $T \times \mathbb{R}^2$ case
considered in {\SSS}3.1.1 of \nocite{RBBSJ06}{Roukema} {et~al.} (2007), 
where the test particle lies along the geodesic joining the three images
of the massive 
particle to one another in one of the spherical spaces, i.e., $S^3/\Gamma$
for an appropriate holonomy group $\Gamma$.
This may help us to understand the full sum from all the adjacent images.
Similarly to Eq.~(2) of \nocite{RBBSJ06}{Roukema} {et~al.} (2007), we can use scalar quantities.
The acceleration to fifth order in $r/\rC$ for the Poincar\'e space is\footnote{The 
expression for the fifth order term is
$             [      68\cos(\pi/5)\sin^4(\pi/5)+
                   240\cos^3(\pi/5) \sin^2(\pi/5)+
                   180\cos^5(\pi/5)
             ]/[15\sin^7(\pi/5)] \;
             (\frac{r}{\rC} )^5. 
$}
\begin{eqnarray}
 \ddot{r}  
&=&
G \frac{m}{\rC^2} 
\left\{
       \left[ \sin \left(\frac{\pi}{5} - \frac{r}{\rC} \right)\right]^{-2}
      - \left[ \sin \left(\frac{\pi}{5} + \frac{r}{\rC} \right)\right]^{-2}
      \right\} \nonumber \\
 & = & G \frac{m}{\rC^2} 
          \left[ 
            \frac{4\cos(\pi/5)}{\sin^3(\pi/5)} \frac{r}{\rC} + \right. 
            \nonumber \\
 &&   \left.  \frac{16\cos(\pi/5)\sin^2(\pi/5)+24\cos^3(\pi/5)}{3\sin^5(\pi/5)}
            \left(\frac{r}{\rC} \right)^3 + \ldots \right] \nonumber \\
&\approx & G \frac{m}{\rC^2} 
          \left[ 15.9 \left(\frac{r}{\rC} \right) +
              81.6  \left(\frac{r}{\rC} \right)^3 + 
              310.4  \left(\frac{r}{\rC} \right)^5 +  \right.
              \nonumber \\
              && \left. \ldots
            \right]. \nonumber \\
&&\rule{\columnwidth}{0ex} \hspace{-8ex}
\label{e-dodec-one-pair}
 (\theequation\addtocounter{equation}{1}) \nonumber  
\end{eqnarray}
This is similar to the flat case. The residual acceleration
is proportional to the displacement to first order, which again is
behaviour similar to that of a cosmological constant: 
\begin{eqnarray}
\ddot{r} \propto r.
\label{e-1d}
\end{eqnarray}
The octahedral space has adjacent images at $(\pi/3)\rC$, so the 
corresponding calculation for a single pair of images is
\begin{eqnarray}
  \ddot{r}  
&=&
G \frac{m}{\rC^2} 
\left\{
       \left[ \sin \left(\frac{\pi}{3} - \frac{r}{\rC} \right)\right]^{-2}
      - \left[ \sin \left(\frac{\pi}{3} + \frac{r}{\rC} \right)\right]^{-2}
      \right\} \nonumber \\
 & = & \frac{16\sqrt{3}}{9} G \frac{m}{\rC^2} 
          \left[ 
           \frac{r}{\rC} + 
            2 \left(\frac{r}{\rC} \right)^3 + 
            \frac{14}{5} \left(\frac{r}{\rC} \right)^5 + 
            \ldots \right] \nonumber \\
&\approx & G \frac{m}{\rC^2} 
          \left[ 3.1 \left(\frac{r}{\rC} \right) +
              6.2  \left(\frac{r}{\rC} \right)^3 + 
              8.6  \left(\frac{r}{\rC} \right)^5 + 
               \ldots
            \right].  \nonumber \\
&&\rule{\columnwidth}{0ex} \hspace{-8ex}
\label{e-octa-one-pair}
(\theequation\addtocounter{equation}{1}) \nonumber  
\end{eqnarray}
Equation~(\ref{e-octa-one-pair}) is also valid for the adjacent
images at the eight truncated corners of the fundamental domain in a 
truncated cube space, which are at $(\pi/3)\rC$. The other six adjacent 
images in the truncated cube space are at $(\pi/4)\rC$, giving
\begin{eqnarray}
  \ddot{r} 
&=&
G \frac{m}{\rC^2} 
\left\{
       \left[ \sin \left(\frac{\pi}{4} - \frac{r}{\rC} \right)\right]^{-2}
      - \left[ \sin \left(\frac{\pi}{4} + \frac{r}{\rC} \right)\right]^{-2}
      \right\} \nonumber \\
 & = & 8 G \frac{m}{\rC^2} 
          \left[ 
           \frac{r}{\rC} + 
           \frac{10}{3} \left(\frac{r}{\rC} \right)^3 + 
            \frac{122}{15} \left(\frac{r}{\rC} \right)^5 + 
            \ldots \right] \nonumber \\
&\approx & G \frac{m}{\rC^2} 
          \left[ 8 \left(\frac{r}{\rC} \right) +
              26.7  \left(\frac{r}{\rC} \right)^3 + 
              65.1  \left(\frac{r}{\rC} \right)^5 + 
               \ldots
            \right]. \nonumber \\
&&\rule{\columnwidth}{0ex} \hspace{-9ex}
\label{e-tcube-one-pair}
(\theequation\addtocounter{equation}{1}) \nonumber  
\end{eqnarray}

Each of these calculations is for just one pair of topological images adjacent
to the ``local'' copy of the massive particle, lying in opposite directions. 
In the flat case, 
this can be thought of as a $T \times \mathbb{R}^2$ model, 
which is a valid 3-manifold.
However, the spherical manifolds are not vector spaces, so 
a similar interpretation of the single-pair calculation is not possible.

%

On the other hand, Eqs~(\ref{e-dodec-one-pair}),
(\ref{e-octa-one-pair}), and (\ref{e-tcube-one-pair}) 
hint at the form that numerical
estimates for the sum of the weak-limit residual gravitational effect from {\em
  all} the adjacent images may take for a given spherical
manifold. For $\PDSmath$, the full set of
adjacent images of the massive particle consists of six pairs of
images. For a test particle displaced slightly in a random direction
from the massive particle, the two images in a pair will be seen in
nearly, although not exactly, opposite directions in the tangent 3-space
at the test particle's location at $\mathbf{r} \in S^3$. 
The modulus of the (vector) residual
acceleration induced by the nearly opposite pair should not be very
different from the expression given in Eq.~(\ref{e-dodec-one-pair}),
although its expression using elementary algebra might appear complicated.

This
suggests that the scalar amplitude of the vector sum of all twelve
accelerations is likely to involve terms of first, third, and fifth
order in $r/\rC$, with coefficients of the order of magnitude of those
in Eq.~(\ref{e-dodec-one-pair}). However, this argument is not exact.
A test particle displaced
from the massive particle in an arbitrary direction does not, in general, lie
along the great circle defined by a given pair of opposite images, and
can at most lie along only one of the great circles defined by the six
pairs of opposite images. Hence, terms with even powers of $r/\rC$
could also appear in the Taylor expansion.

These single-pair calculations might also be of interest for 
ill-proportioned \nocite{WeeksWellProp04}({Weeks} {et~al.} 2004)
positively curved spaces, e.g., the lens spaces 
L($p,q$), with $p,q \in \mathbb{Z}$ relatively prime, 
where $p \gg 1$ \nocite{GausSph01}(e.g. \SSS4,  {Gausmann} {et~al.} 2001). 
Since these spaces are not globally
homogeneous, derivations similar to those in 
Eqs~(\ref{e-dodec-one-pair}), (\ref{e-octa-one-pair}), and (\ref{e-tcube-one-pair}) would
be strictly valid only for points lying along the symmetry axis
joining the centres of the two faces of the fundamental dihedron (lens),
i.e., where adjacent topological images 
are separated by a spatial geodesic of length $2\pi \rC/p$. 
This direction would therefore be
unstable to a linear-term acceleration effect that would tend
to expand it faster than other directions. 
This is qualitatively similar to the effect found in {\SSS}3 of 
\nocite{RBBSJ06}{Roukema} {et~al.} (2007), according to which the residual acceleration would
tend to equalise the three fundamental lengths of a ${T}^3$ model
of slightly unequal fundamental lengths. 

\foctaAthree

\ftcubeAthree

\fdodecAthree


\subsection{Well-proportioned spherical spaces: $S^3/T^*$, $S^3/O^*$, and $\PDSmath$ }
\label{s-results-wellprop}

As described in Sections~\ref{s-meth-resgrav} and 
\ref{s-meth-numerical}, for a test particle at
$\mathbf{r}$ in $S^3/T^*$, $S^3/O^*$, or $\PDSmath$, let us set $\rhor
= 1.1 (\pi/3) \rC, $ $\rhor = 1.1 (\pi/3) \rC, $ or $\rhor = 1.1
(\pi/5) \rC $, respectively. Equation~(\ref{e-residgrav-S3-Gamma}) is evaluated
numerically for 200 randomly (logarithmically) distributed test
particles at distances of up to 250{\hMpc} from the massive particle.
Figures \ref{f-octa-a3}, \ref{f-tcube-a3}, and \ref{f-dodec-a3} show
the residual accelerations, scaled by $(r/\rC)^3$ so that they should
be approximately constant 
if the dominating term in $r/\rC$ is the third
order term, i.e.,
\begin{equation}
a_3(r) \equiv \frac{\ddot{r}}{Gm/\rC^2} \left(\frac{r}{\rC}\right)^{-3}
\label{e-define-a3-S3}
\end{equation}
is shown.

\fdodecAthreeperp

\tcoeff

It is clear that both the octahedral and
truncated cube spaces have similar Taylor series behaviour to that of $T^3$. The
linear term cancels, but a third-order-dominated residual acceleration
remains.  On the other hand, it is clear from Fig.~\ref{f-dodec-a3} that the
Poincar\'e dodecahedral space has a residual acceleration that is much weaker
than those of the other three spaces, and that this is poorly
modelled as a third order term in $r/\rC$. 
Instead, the constant slope of the relation in Fig.~\ref{f-dodec-a3} 
strongly suggests
that the residual acceleration for the Poincar\'e space is
dominated by a {\em fifth} order term, i.e.,
\begin{equation}
a_5(r) \equiv \frac{\ddot{r}}{Gm/\rC^2} \left(\frac{r}{\rC}\right)^{-5}
\label{e-define-a5-S3}
\end{equation}
is approximately constant and terms lower than the fifth order cancel.

The remaining orthogonal component of the residual acceleration is
of a similar order of magnitude to that of the radial component. 
Figure~\ref{f-dodec-a3-perp} 
shows that provided that the component is calculated with at least 
70-bit precision in the significand, it is dominated by the $(r/\rC)^5$ term.

The previous
figures were calculated using 70-bit significand multi-precision
floating-point operations. 
The effect of
reducing the precision to 50 or 60 bits is clear
in Fig.~\ref{f-dodec-a3-perp}, i.e.,
for the orthogonal component of the residual
acceleration in the Poincar\'e space. 
Noise created by the precision limit enters the 
calculation, for example, 
when converting from a double-precision position in $\mathbb{R}^4$ of
the test particle to a multi-precision position. Apart from special cases,
the finite precision
representation of a typical, pseudo-random particle position in $\mathbb{R}^4$ will
place the particle at a position slightly offset 
from the physical 3-surface $S^3$, because
of the limited precision. If
the particle is not located exactly on $S^3$, then
this induces an acceleration that is orthogonal to the tangent 3-plane at 
$\mathbf{r}$. 
The numerical noise component appears on the left of 
Fig.~\ref{f-dodec-a3-perp} as a constant
value of $\ddot{r}$, i.e., $a_3(r) \propto (r/\rC)^{-3}$.

\subsubsection{Analytical estimate for the Poincar\'e space}
\label{s-results-analytical}

Analytical evaluation of 
Eq.~(\ref{e-residgrav-S3-Gamma}) to fifth order for the Poincar\'e
dodecahedral space using a 
{computer algebra 
system\footnote{{The script is available online 
at \url{http://adjani.astro.umk.pl/GPL/dodec/PDS_residual}. Version 1.0 
was used in this paper.}}}
confirms the numerical estimates shown in Figs.~\ref{f-dodec-a3} and \ref{f-dodec-a3-perp}. 
The highest order residual acceleration from the adjacent topological images 
is
\vbox{
\begin{eqnarray}
\ddot{\mathbf{r}} &=&
\frac{
12 \sqrt{2} \left(297 \sqrt{5} + 655\right)
}{125 \sqrt{5-\sqrt{5}}}
\left(\frac{r}{\rC}\right)^5
 \nonumber \\  && 
\Big\{
\big[70\,y^4
+(42\,\sqrt{5}+70)\,x^2\,y^2
-(14\,\sqrt{5}+70)\,y^2    \nonumber \\  && 
+(21\,\sqrt{5}-7)\,x^4
-28\,\sqrt{5}\,x^2
+7\,\sqrt{5}+5\big] \;x,    \nonumber \\  && 
\big[70\, z^4
+(42\,\sqrt{5}+70)\,y^2\,z^2
-(14\,\sqrt{5}+70)\,z^2      \nonumber \\  && 
+(21 \,\sqrt{5}-7)\,y^4
-28\,\sqrt{5}\,y^2
+7\,\sqrt{5}+5\big] \;y,     \nonumber \\  && 
\big[
70\,x^4
+(42\,\sqrt{5}+70) \,x^2\,z^2
-(14\,\sqrt{5}+70)\,x^2      \nonumber \\  && 
+(21\,\sqrt{5}-7)\,z^4
-28\,\sqrt{5}\,z^2
+7\,\sqrt{5}+5\big] \;z ,     \nonumber \\  && 
0
\Big\} , 
\label{e-residgrav-Poincare-exact}
\end{eqnarray}
}
where the massive particle is at $(0,0,0,1)$, the
test particle displaced from it in an arbitrary direction by 
a small amount is located at
$[\sin(r/\rC) x, \sin(r/\rC) y , \sin(r/\rC) z, \cos(r/\rC)]$, 
and $x^2+y^2+z^2=1$.
The highest order term in the fourth component of this
expression is a sixth order term in powers of $r/\rC$.
The fifth order exact expression can 
be written with approximate decimal coefficients as
\vbox{
\begin{eqnarray}
\ddot{\mathbf{r}} &\approx&
%
%
%
(r/\rC)^5 \;
[( 
4304.3 x^4
+7540.5 y^4
+17657.2 x^2 y^2
  \nonumber \\ &&
-6744.5 x^2 
-10912.8 y^2 
+ 2224.7
) x 
, \nonumber \\ 
&&(
4304.3 y^4
+7540.5 z^4
+17657.2 y^2 z^2
  \nonumber \\ &&
-6744.5 y^2 
-10912.8 z^2 
+ 2224.7
) y , \nonumber \\
&&
(
4304.3 z^4
+7540.5 x^4
+17657.2 z^2 x^2
  \nonumber \\ &&
-6744.5 z^2 
-10912.8 x^2 
+ 2224.7
) z, \nonumber \\
&&
0
]  .
\label{e-residgrav-Poincare}
\end{eqnarray}
}

The radial and orthogonal components of the residual acceleration 
can now be calculated as follows.
Firstly, a small numerical value $r/\rC = 10^{-3}$ 
is chosen in order to evaluate
the unit tangent vector at $\mathbf{r}$ pointing towards the
massive particle. The sign of the latter is reversed in order to point away from
the massive particle.
Secondly, pseudo-random numerical values for $x,y,z$ 
are substituted
into the full vector residual acceleration 
[Eq.~(\ref{e-residgrav-Poincare-exact}) or (\ref{e-residgrav-Poincare})]
and the scalar product
with the unit tangent vector is taken in order to obtain the radial component
of the residual acceleration. This is subtracted from the full residual to
obtain the orthogonal component.
%

\subsubsection{Statistical description}
\label{s-results-stats}

For an isotropic distribution of the directions of displacement
of the test particle, Table~\ref{t-coeff} lists characteristic 
statistics of the radial and orthogonal components of the 
residual acceleration for the four well-proportioned spaces, primarily
calculated from direct numerical estimates
using Eq.~(\ref{e-residgrav-S3-Gamma}).
By construction, the orthogonal component is necessarily non-negative. 
Since estimation of the parameters of
the Poincar\'e space residual acceleration is sensitive
to numerical precision limits, Table~\ref{t-coeff} also lists parameters
for the Poincar\'e space radial component estimated 
using Eq.~(\ref{e-residgrav-Poincare-exact}) and the unit tangent vector
described above.
Within the uncertainties indicated by the standard error in the mean
$\sigma_{\left<a_i\right>}$, the two estimates agree.

\section{Discussion} \label{s-disc}

The results shown above are quite sensitive to small errors
in the positions of the adjacent topological images. This 
provides a useful check of the calculations, since a small 
arbitrary error is most likely to yield a stronger
residual acceleration than a weaker one.
Increasing or decreasing the distances from the massive particle to the two
members of one of the opposite pairs of topological images by $0.1\%$ 
is sufficient to destabilise the nearly perfect equilibrium defined by the full
set of adjacent topological images for a given space. The residual acceleration
reverts to being dominated by $(r/\rC)^1$ rather than $(r/\rC)^3$
or $(r/\rC)^5$.

This sensitivity to small changes in the distances to topological
images is physically interesting.
As found in {\SSS}3 of 
\nocite{RBBSJ06}{Roukema} {et~al.} (2007), the residual acceleration in a $T^3$ model
of slightly unequal fundamental lengths is dominated by a linear
term in $r/L_a$, which tends to equalise the three fundamental lengths. 
The same relation applies for the spherical models.
These are very well balanced in the sense that the linear term
cancels, and in the case of the Poincar\'e space, the cubic term
cancels too, provided that these spaces are perfectly homogeneous. 
A small decrease in the injectivity diameter in one direction implies
stronger acceleration for a test particle, favouring a return towards
perfect isotropy in the fundamental lengths. Conversely, a small 
{increase} in the injectivity diameter implies
{weaker} acceleration, again favouring a return towards isotropy. 
The residual acceleration tends
to encourage the space to return to perfect 
``residual acceleration equilibrium'', 
in which the residual acceleration disappears down to 
the third or fifth order.

Clearly, these effects are negligible in the present epoch at
the level of observational detectability for the next few decades.
Moreover, given the present empirical interest in the Poincar\'e space,
the prediction of the perfectly isotropic model would be an {\em absence} of a residual
acceleration effect to a much higher observational accuracy than that
for the effects predicted by the other three well-proportioned
models, which in turn would be more difficult to detect than the residual
acceleration effect from ill-proportioned models. 

However, the Universe is certainly inhomogeneous.
If evidence continues to accumulate for the Poincar\'e space
model with successively more accurate estimates of the model's
astronomical coordinates, following the estimates made 
in \nocite{RBSG08,RBG08}{Roukema} {et~al.} (2008a, 2008b), then several observational approaches
could be used to test the predictions of the residual acceleration 
effect. With a sufficient level of precision, matched circles 
and/or annuli on the sky should yield slightly different fundamental
lengths along the six axes. The residual acceleration effect should
consist of a positive or negative extra acceleration for shorter or longer
fundamental lengths, respectively. Surveys of tracers of large-scale structure
lying along the different geodesics to topological images of, for
example, the Virgo cluster, could also be used to estimate which directions 
should have slightly longer than average geodesics and which should have
slightly shorter than average geodesics. All three estimates should
agree with one another, provided that no other sources of
random and systematic error interfere.
Hence, the residual acceleration effect potentially offers a 
physical mechanism for testing
a multiply-connected model of comoving space, as an alternative to
relying on geometrical effects (individual or statistical
identification of multiply imaged objects or regions of space). 

\tcoefftcube

Another interesting question is whether or not the better 
``balancing'' of the Poincar\'e space could have had any role
in the evolution of a preferred topology in the early Universe,
especially during the quantum epoch. Does the Poincar\'e space 
occupy a dynamically more stable state than the other well-proportioned
spaces, which in turn are more stable than ill-proportioned spaces?
This would most likely require significant inhomogeneities in 
the very early Universe, which themselves might have been subsequently removed
by the residual gravity acceleration effect itself.

Apart from the possible physical consequences of this result, it may
be interesting to ask if these results are reasonable from an
intuitive mathematical point of view.  It is useful to consider this
in terms of pairs of opposite images, since if the test particle lies
along the geodesic joining the local copy and the two opposite distant
copies of the massive particle, then the single pair yields a
linear-dominated residual acceleration.  In order for the linear (or
higher order) terms in the residual acceleration to cancel, the
acceleration components that are approximately orthogonal to the
geodesic joining the test particle to the two members of a given opposite
pair need to be able to cancel the approximately radial components of
residual accelerations in other pairs of images. In the $T^3$ case, three 
nearly orthogonal pairs
are sufficient for the linear terms to completely cancel in this way
(provided that the fundamental domain is perfectly regular).
Given that $S^3/T^*$, $S^3/O^*$, and $S^3/I^*$ have four, seven, and
six opposite pairs, respectively, it is reasonable to imagine that the radial
components of the linear residual-acceleration term of a given
pair are cancelled by the orthogonal components of the other pairs,
since there are many of them, distributed symmetrically.

It is also possible to imagine that with a high enough number of
opposite pairs and sufficient symmetry, a higher order term such as
the third order might also disappear, by means of fine balancing between
the different directions.  The truncated cube space $S^3/O^*$ and the
Poincar\'e space $S^3/I^*$ have the highest numbers of opposite pairs,
or equivalently, the highest numbers of pairs of faces of their
fundamental domains.  However, the truncated cube cannot be a regular
polyhedron; its faces are six truncated squares and eight
triangles. All the faces of the fundamental domain of the 
Poincar\'e dodecahedral space have the same shape: a pentagon.
In this sense, it is more symmetrical than the truncated cube space.
So requiring both a fundamental domain with a high number of faces and a
high level of symmetry would seem to provide a qualitative pair
of conditions to explain why the Poincar\'e space is more
finely balanced than the other well-proportioned spaces.

However, could either the subset of
topological images corresponding to only the octagonal faces of
the fundamental domain of the truncated cube, or the subset
corresponding to only the triangular faces, nevertheless cancel
to fifth order, and become hidden because the other subset
reintroduces a third order term?  A calculation similar to those developed 
above shows
that both subsets separately retain the third order as the
dominant order.  Table~\ref{t-coeff-tcube} gives statistical 
characteristics of the third
order coefficients for the two subsets.
So the Poincar\'e space retains its uniqueness in cancelling
down to fifth order.

An aspect of the residual acceleration as expressed in 
Eqs~(\ref{e-residgrav-Poincare-exact}) and
(\ref{e-residgrav-Poincare})
that may initially
seem counterintuitive is the 
permutational symmetry between the $x,y,$ and $z$ components, 
since a dodecahedron is not always thought of as having cubical
symmetry. However, an appropriate orientation of the dodecahedron in
$\mathbb{R}^3$, 
{such as that used in our computer algebra script,}
shows an $(x,y,z)$ symmetry. {The twelve adjacent
topological images
consist of three quadruplets orthogonally projected to $\mathbb{R}^3$,  
$[0, \pm 1/2, \pm (\sqrt{5}-1)/4]$, 
$[\pm (\sqrt{5}-1)/4, 0, \pm 1/2]$, and
$[\pm 1/2, \pm (\sqrt{5}-1)/4, 0]$.
Each of these quadruplets} lies 
in a 2-plane in $\mathbb{R}^3$, 
{i.e., $x=0$, $y=0$, and $z=0$, respectively.}
In this orientation, the ``top'' and ``bottom'' of the dodecahedron
are edges, not faces. These two edges, along with four others, can
be used to inscribe the dodecahedron in a cube (in flat space).

\section{Conclusions} \label{s-conclu}

In \nocite{RBBSJ06}{Roukema} {et~al.} (2007), it was found that a residual gravity acceleration 
effect exists for $T\times \mathbb{R}^2$ but cancels to first order
for a perfectly regular $T^3$ model. Here, it was found that the residual
acceleration in $T^3$ does not completely cancel. In $T^3$ and two
other well-proportioned spaces, the octahedral space $S^3/T^*$ and
the truncated cube space $S^3/O^*$, the residual acceleration exists
as a third order effect in the Taylor expansions in 
$r/L_a$ and $r/\rC$ respectively.
A reasonable upper limit to $r/L_a = r/\rC$ can be set by considering
test particles 
in voids of large-scale structure, i.e., at typically a
few megaparsecs from the most massive nearby object, and an injectivity
diameter for the 3-manifold of comoving space of at least a few gigaparsecs,
i.e., $r/L_a = r/\rC \sim 10^{-3}$.
Hence, at the present epoch, these three well-proportioned spaces are about a
million times better balanced by this dynamical criterion than
ill-proportioned spaces. 

The Poincar\'e dodecahedral space $S^3/I^*$,
presently a candidate 3-manifold for comoving space favoured by
several observational analyses, has been found to be even more exceptional. Its
residual acceleration is dominated by the fifth order term of amplitude
$\sim \pm 300 (Gm/\rC^2) (r/\rC)^5$ 
[Figs.~\ref{f-dodec-a3}, \ref{f-dodec-a3-perp},
Table~\ref{t-coeff}, 
Eqs~(\ref{e-residgrav-Poincare-exact}), (\ref{e-residgrav-Poincare})].
This makes it about ten thousand times better
balanced than the other three well-proportioned spaces, i.e., about
$10^{10}$ times better balanced than ill-proportioned spaces.
Moreover, perturbations to this equilibrium
favour a return to the equilibrium state.
Are these clues towards a theory of cosmic topology?

\begin{acknowledgements}

Thank you to Zbigniew Buli\'nski and Bartosz Lew for helpful discussion
{and to an anonymous referee for useful suggestions.}
%
%
Use was made 
of the
Centre de Donn\'ees astronomiques de Strasbourg 
(\url{http://cdsads.u-strasbg.fr}), 
the computer algebra program {\sc maxima},
the GNU {\sc Octave} command-line, high-level numerical computation software 
(\url{http://www.gnu.org/software/octave}),
the GNU multi-precision library (GMP) and the MPFR library,
the GNU Scientific Library (GSL), 
and the GNU {\sc plotutils} plotting package.

\end{acknowledgements}

\subm{ \clearpage }

\nice{
%

}


\newcommand\astroph[1]{}

{
\appendix
\section{Residual gravity in the Poincar\'e space: computer algebra script}
\label{a-maxima}

The following script evaluates the residual acceleration on a
test particle near a massive particle due to the adjacent
topological images of the massive particle, as given in
Eq.~(\ref{e-residgrav-S3-Gamma}), for the Poincar\'e
dodecahedral space. The script is written using
the syntax of the computer algebra system {\sc maxima}, version 5.10.0.

\begin{verbatim}
/*
# topograv - numerical tests of residual gravity
# force induced by spatial topology
#
# Copyright (C) 2009 Boud Roukema
#
# This program is free software; you can
# redistribute it and/or modify it under the
# terms of the GNU General Public License as
# published by the Free Software Foundation;
# either version 2, or (at your option) any
# later version.
#
# This program is distributed in the hope that
# it will be useful, but WITHOUT ANY WARRANTY;
# without even the implied warranty of
# MERCHANTABILITY or FITNESS FOR A PARTICULAR
# PURPOSE.  See the GNU General Public License
# for more details.
#
# You should have received a copy of the GNU
# General Public License along with this
# program; if not, see:
#       http://www.gnu.org/licenses/gpl.html
*/

/* avoid imaginary sqrt ambiguities */
radexpand: false;

normalise_v4(x) := [ x[1]/sqrt(x.x),
 x[2]/sqrt(x.x), x[3]/sqrt(x.x), 
 x[4]/sqrt(x.x) ];


/* normalised tangent at 4-vector a pointing
   towards 4-vec b */ 
tangent_S3_at_a(a, b) :=
 normalise_v4(b - ((a . b) * a));

/* spherical-newtonian acceleration at a from
   image at b */ 
accel_at_a(a,b) := tangent_S3_at_a(a, b) /
 (1- (a . b)^2);

/* zeroth image of massive particle is at
   [0,0,0,1] */

/* number related to the Golden Ratio */
phi_m14 : (sqrt(5)-1)/4 ;
phi_p14 : (sqrt(5)+1)/4 ; /* = cos(%pi/5) */

/* 12 adjacent images of massive particle;
   modulus must be 1 since the physical space 
   is S^3, not R^4 */
image1 : [ 0 ,  1/2, -phi_m14,  phi_p14 ];
image2 : [ 0 , -1/2, -phi_m14,  phi_p14 ];
image3 : [ -1/2,   phi_m14, 0,  phi_p14 ];
image4 : [  1/2,   phi_m14, 0,  phi_p14 ];
image5 : [ -phi_m14,  0, -1/2,  phi_p14 ];
image6 : [ -phi_m14,  0,  1/2,  phi_p14 ];

image7 : [ 0 , -1/2,  phi_m14,  phi_p14 ];
image8 : [ 0 ,  1/2,  phi_m14,  phi_p14 ];
image9 : [  1/2,  -phi_m14, 0,  phi_p14 ];
image10 : [ -1/2, -phi_m14, 0,  phi_p14 ];
image11 : [ phi_m14,  0,  1/2,  phi_p14 ];
image12 : [ phi_m14,  0, -1/2,  phi_p14 ];



/* sum of accelerations */
accel_full(vec_m) := 
	     accel_at_a(vec_m,image1) +
	     accel_at_a(vec_m,image2) +
	     accel_at_a(vec_m,image3) +
	     accel_at_a(vec_m,image4) +
	     accel_at_a(vec_m,image5) +
	     accel_at_a(vec_m,image6) +
	     accel_at_a(vec_m,image7) +
	     accel_at_a(vec_m,image8) +
	     accel_at_a(vec_m,image9) +
	     accel_at_a(vec_m,image10) +
	     accel_at_a(vec_m,image11) +
	     accel_at_a(vec_m,image12);

/* test particle near to zeroth image of
   massive particle, in arbitrary direction
  (x,y,z); modulus must be 1 since the physical
   space is S^3, not R^4 */
vec_m : [ sin(rr)*x, sin(rr)*y , sin(rr)*z,
 cos(rr) ];


/* Taylor expand to fifth order */
tay : taylor(accel_full(vec_m),rr,0,5);   

/* simplify x, y components */
tay_x : gfactor(rectform(expand(tay[1]))), 
    z : sqrt(1-x^2-y^2); 
tay_y : gfactor(rectform(expand(tay[2]))), 
    x : sqrt(1-y^2-z^2); 
/* simplify  z  component; to retain symmetry,
   reverse alphabetical order of (x,z)
   by temporarily substituting (b,a); 
*/
subst(a,z,subst(b,x,tay[3]));
gfactor(rectform(expand(%))), 
    y : sqrt(1-a^2-b^2);
tay_z : subst(z,a,subst(x,b,%));
/* simplify  w  component */
tay_w : expand(tay[4]), 
    x : sqrt(1-y^2-z^2);

/* still fully algebraic */
residual_fully_alg : 
     [tay_x, tay_y, tay_z, tay_w ];

/* 212-bit precision for big floats 1b0 etc. */ 
fpprec : 64; ratepsilon : 1b-64;

/* numerical coefficients in front of x,y,z */
residual_xyz_alg : residual_fully_alg, bfloat;

/* put some weakly pseudo-random numbers into
   x,y,z, while requiring the R^3 modulus to 
   be 1 so that vec_m is on S^3; */
x: 2b0*random(1.0)-1b0;
theta: 2b0*%pi *random(1.0);
y: sin(acos(x)) * cos(theta), bfloat;
z: sin(acos(x)) * sin(theta), bfloat;

residual : residual_xyz_alg, bfloat;


/* components */
radial_n: -tangent_S3_at_a(vec_m, [0, 0, 0, 1]), 
                 rr : 1b-3; 
residual_radial: expand(residual . radial_n);

ortho: expand(residual - 
                 residual_radial * radial_n);
residual_ortho: expand(sqrt(ortho.ortho));

/* result summary */
print("vector residual (coeff, x,y,z algeb.)"); 
residual_fully_alg ; 
print("vector residual (x,y,z algebraic)"); 
residual_xyz_alg;
print("vector residual (x,y,z numeric)"); 
residual;
print("radial residual (x,y,z numeric)"); 
residual_radial;
print("orthogonal residual (x,y,z numeric)"); 
residual_ortho;
\end{verbatim}
}   


\end{document}